\documentstyle[11pt,dunk2001_asp,twoside,psfig]{article}
\def\edcomment#1{\iffalse\marginpar{\raggedright\sl#1\/}\else\relax\fi}
\reversemarginpar

\begin{document}
\title{Automating the Synthetic Field Method:\\
Application to Sextans A}
\author{Benne W. Holwerda, Ronald J. Allen}
\affil{Space Telescope Science Institute, 3700 San Martin Drive,
Baltimore, MD 21218, USA}
\author{and Pieter C. van der Kruit}
\affil{Kapteyn Institute, Landleven 12, 9747 AD Groningen, the Netherlands}

\begin{abstract}
We have automated the ``Synthetic Field Method'' developed by Gonzalez et al.
(1998) and used it to measure the opacity of the ISM in the Local
Group dwarf galaxy Sextans A by using the changes in counts
of background galaxies seen through the foreground system.
The Sextans A results are consistent with the observational relation 
found by Cuillandre et al. (2001) between dust opacity and HI column 
density in the outer parts of M31.
\end{abstract}

\section{Introduction}

To observe the opacity and distribution of dust independent of theoretical 
models for the light or dust distribution, the dust needs to be backlighted
 by a more distant source. The group led by Keel and White used an occulted 
galaxy for this purpose and their studies (Domingue et al. (1999, 2000), 
Keel et al. (2001) and White et al. (1992, 2000)) showed insight into the 
fine structure of the dust in a few galaxies.

Gonzalez et al. (1998) developed a method to use the distribution of field 
galaxies which they called the ``Synthetic Field Method'' (SFM); it is not 
restricted to the case of nearby galaxy pairs and it can supply 
values for the dust opacities of galaxies in the disks of spirals 
and irregulars in general.

\section{The Synthetic Field Method}

The method consists of comparing the numbers of the background 
galaxies found in Wide Field Camera images of nearby spiral and irregular 
galaxies with those of simulations. These simulations are a Hubble Deep 
Field image, extincted by a certain opacity added to the original data. (see figure 1.) 
By adding the extincted HDF frame to the original data, all the effects of 
clustering and confusion by the foreground galaxy are calibrated. All the 
three HDF frames are used in seperate simulations to account as much as 
possible for clustering biases; however, clustering of the field galaxies 
remains the main uncertainty. Visual identification of the 
background galaxies in both the data and the simulations is very time 
consuming. Therefore we have searched for ways to automate the procedure.

\begin{figure}
\centerline{\vbox{
\psfig{figure=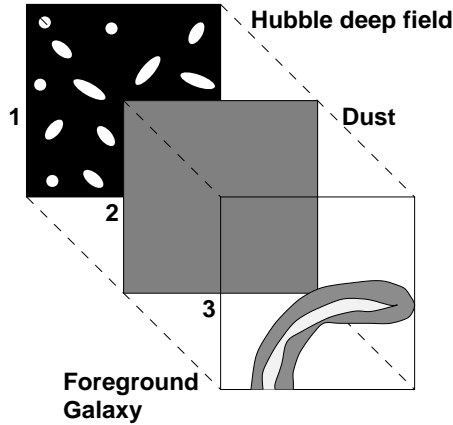,width=6cm,angle=0}
}}
\caption{A conceptual view of a simulation: (1) a Hubble deep field frame, (2) artificial dust, and (3) the foreground galaxy data.}
\end{figure}

\subsection{Automating the SFM}

Sextans A was the test case for automating the SFM together with NGC4536.
In both cases results from manual galaxy counts were available. 
The automation works briefly as follows:
a modified Source Extractor v2.2.2 (Bertin \& Arnouts 1996)
provides catalogues of objects with a variety of structural parameters.
(Source Extractor's neural network classification, concentration, contrast,
 flux radius, area, ellipticity, color and asymmetry).
\footnote{Concentration, contrast and asymmetry were defined as in 
Abrahams (1993).}

Candidate objects are selected based on a score which is determined
by a series of criteria for the structural parameters;
values typical for background galaxies earn an object a higher score while 
those typical for stars and globular clusters result in a lower score.
The advantages of the system are that objects does not need to meet all 
the criteria to be selected, and criteria can involve combinations of 
parameters.

The selected objects in the original data were checked visually
to exclude diffraction spikes and blended stars. The selected objects 
in the simulations were automatically checked for the data-objects, 
thus leaving only simulated galaxies. 

\begin{table}
\caption{Total exposure times and number of exposures in brackets 
of the data on Sextans A. The exposures were 'drizzled' using the 
program from Fruchter et al. (1997). }
\begin{tabular}{l | c c c }
Position	& Exposure Time (s) 	&  & final pixelscale \\
\hline
		& I			& V  & \\
Sextans A North & 38400	(32)& 19200 (16) & 0.05''  \\
Sextans A South & 1800 (3)	& 1800 (3) & 0.05'' \\
\end{tabular}

\end{table}

\section{Sextans A data}

Sextans A provided an excellent opportunity to test the automated  
Synthetic Field Method with data on most of the galaxy with varying
exposure times but with the two filters (V (F555W) and I (F814W))
 used by Gonzalez et al. (1998) (tabel 1).

HI maps from Skillmann et al. (1988; figure 3) showed that the HST fields
 covered a range of HI column densities; figure 3) This allows us to 
compare dust to gas ratios similar to the approach of 
Cuillandre et al. (1999) in the outskirts of M31.

\section{Results}

The results for the WF3 and 4 are consistent with 0 magnitude opacity 
and WF2 with 0.5 magnitude. (see figure 2) 
However the error in the number of background galaxies due to their 
clustering is such that all can be considered consistent with no dust opacity.
As can be seen from figure 4 the determined opacities for Sextans A all 
correspond to the relation Cuillandre et al. (2001) found experimentally in 
Sextans A. 

\begin{figure}
\centerline{
\hbox{
\psfig{figure=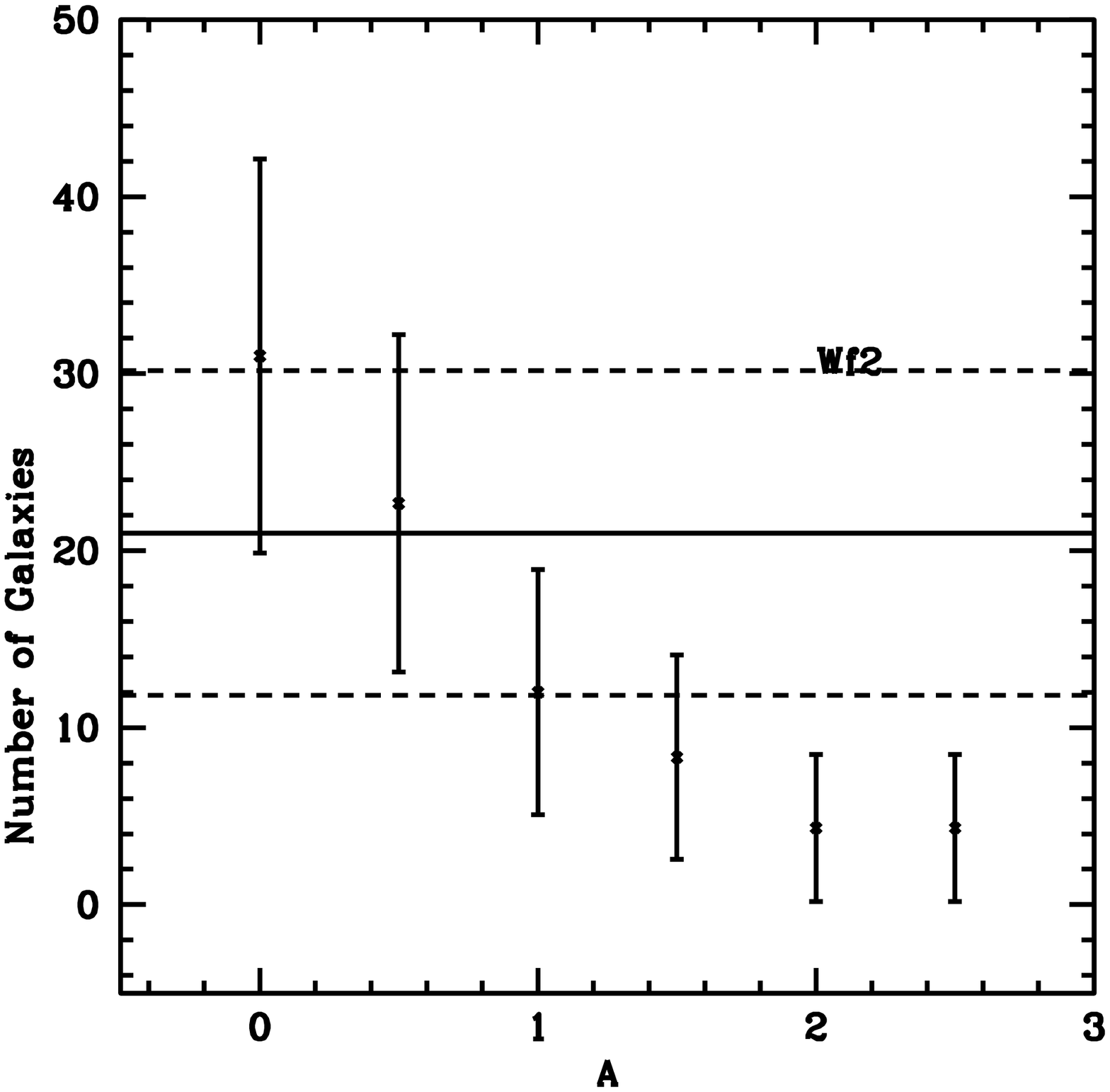,width=4cm,angle=0}
\psfig{figure=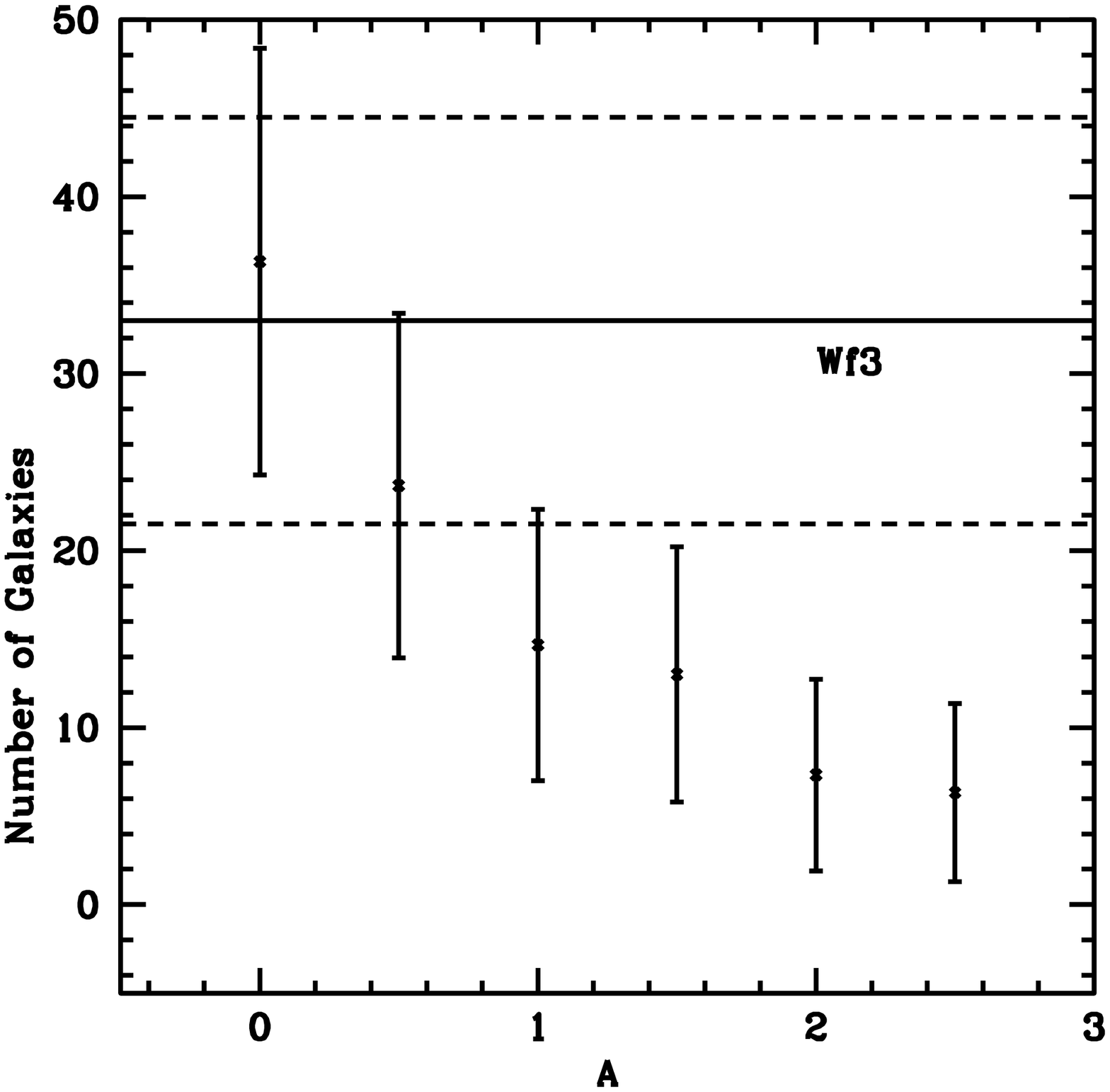,width=4cm,angle=0}
\psfig{figure=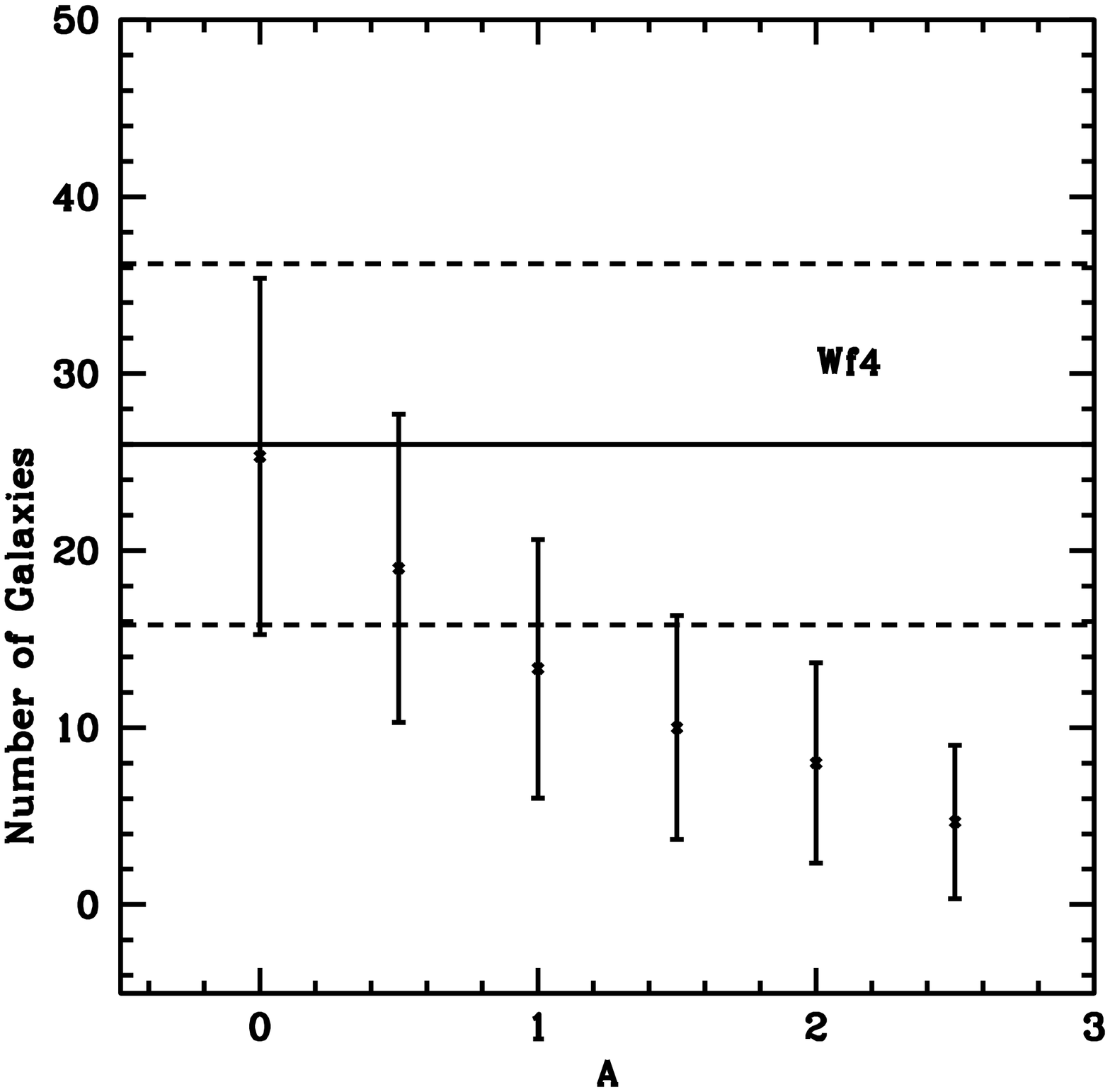,width=4cm,angle=0}
}}
\caption{Plots of the number of galaxies found in simulations (points) 
and the real data (solid line, error indicated with dashed line) 
for all three Wide Field CCD's. The simulations are done for a range of possible extinction values A. The errors are taken to be twice the Poisson 
errors to account for clustering error.
Gonzalez et al. (1998 and in preparation) provide a more extensive 
discussion of the error estimate.}
\end{figure}

\begin{figure*}
\centerline{
\hbox{
\psfig{figure=./holwerda_fig5.ps,width=7cm,angle=0}
\psfig{figure=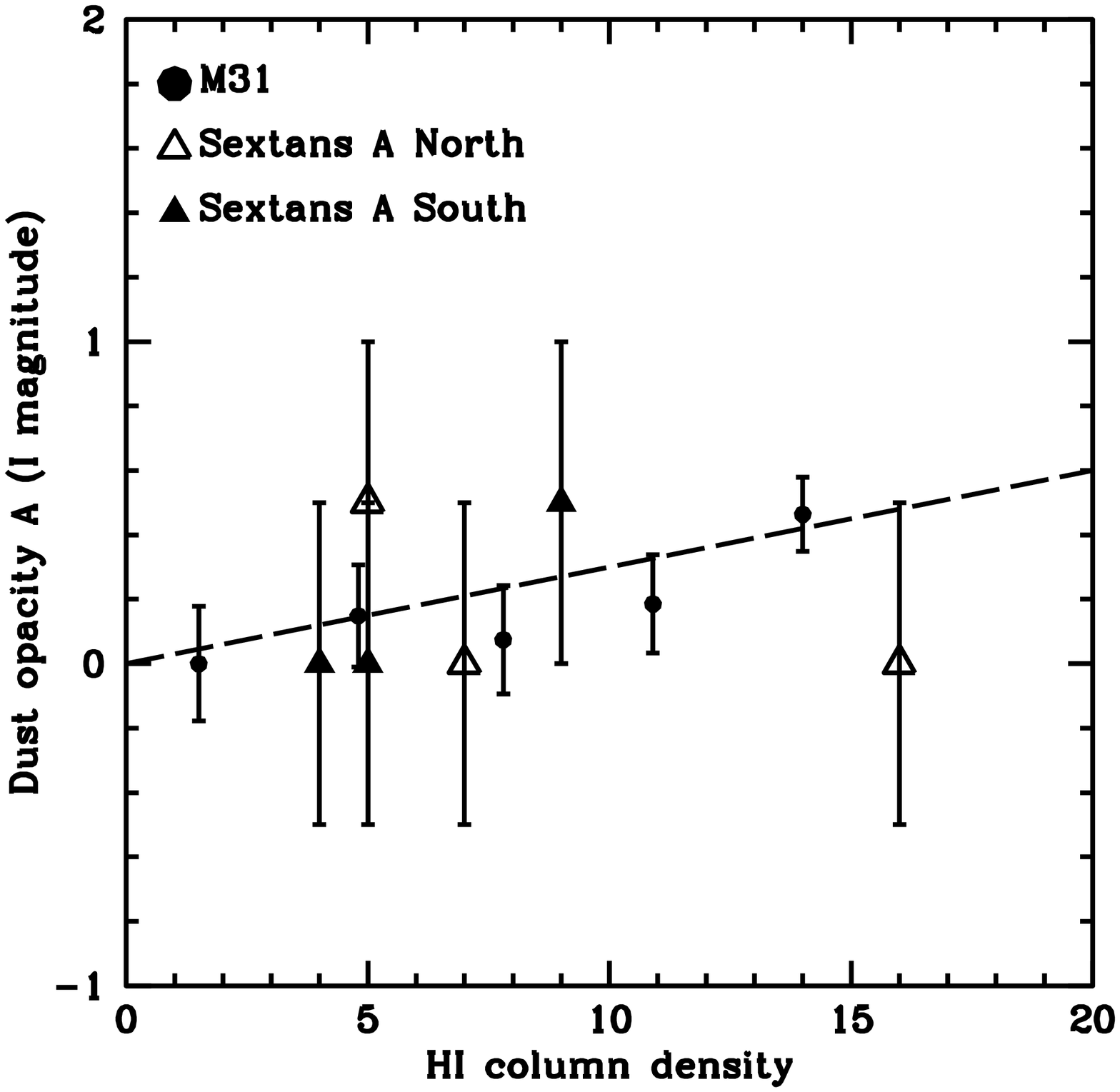,width=6cm,angle=0}
}}
\caption{The HI map of Sextans A from Skillmann et al. (1988) with the position of the 
WFPC fields (north solid, south dashed) overlayed. Contours are 
drawn at 2.5,5,10,15,20,25 and 30 $10^{20}$ atoms per cm$^2$ }
\caption{Opacities in I magnitude against HI column densities in $10^{20}$ 
atoms per cm$^2$. The results from Sextans A North (open triangles and 
South (filled triangles ) follows the relation found by Cuillandre et al.
 (2001) (dashed line) from their ground based results (filled dots). 
Errors scale inversely with area due to field galaxy clustering.} 
\end{figure*}

\section{Discussion}

We conclude first that the SFM can be applied with almost the same 
accuracy on fields with widely different exposure times. The 
limitations come from background galaxy clustering and confusion, 
not sensitivity. 
Secondly, the SFM can be automated to a satisfactory degree.
As the clustering is the main uncertainty, applying the SFM to larger areas 
with similar properties (HI column density, inside or outside the spiral arm 
or at a certain radius) will lower these errors and give a general picture of 
dust distribution in irregular and spiral galaxies.\\

We intend now to carry out a survey using 
the SFM of suitable WFPC data currently in the HST archive.


\begin{references}
\reference{Bertin, E. and Arnouts, S., 1996, AAPS, 117, 393}
\reference{Cuillandre, J., Lequeux, J., Allen, R.J., Mellier, Y. and Bertin, E., 2001,\\ ApJ, 554, 190}
\reference{Domingue, D.L., Keel, W.C., Ryder, S.D. and White, R.E., 1999, AJ, 118, 1542}
\reference{Domingue, D.L., Keel, W.C. and White, R.E., 2000, ApJ, 545, 171}
\reference{Fruchter, A. and Hook, R.N., Proc. SPIE, 3164, 120}
\reference{Gonz{\' a}lez, R.A., Allen, R.J., Dirsch, B.,
        Ferguson, H.C., Calzetti, D. \\ and Panagia, N.,1998, ApJ, 506, 152}
\reference{Keel, W.C. and White, R.E., 2001, AJ, 121, 1442}
\reference{Skillman, E.D., Terlevich, R., Teuben, P.J. and van Woerden, H., 1988,
\\ AAP, 198,33}
\reference{White, R.E. and Keel, W.~C., 1992, Nat, 359, 129}
\reference{White, R.E., Keel, W.C. and Conselice, C.J., 2000, ApJ, 542, 761}
\end{references}
\end{document}